\documentclass[prb,twocolumn,superscriptaddress]{revtex4-2}
\usepackage{amsfonts}
\usepackage{amssymb}
\usepackage{amsmath}
\usepackage{graphicx}
\usepackage{epstopdf}
\usepackage{dcolumn}
\usepackage{bm}
\usepackage{units}
\usepackage{soul}
\usepackage{multirow}
\usepackage{CJKutf8}
\usepackage{subfigure}
\usepackage{color}%
\usepackage{url}
\usepackage[colorlinks,linkcolor=blue,anchorcolor=blue,citecolor=blue,urlcolor=blue]{hyperref}
\usepackage{array}
\usepackage{ulem}
\usepackage{float}
\begin{document}
\title{Anisotropic magnetoresistance in single cubic crystals: A theory and its verification}
%\title{Universal features of anisotropic %magnetoresistance in single cubic crystals}
\author{Yu Miao}
\thanks{Yu Miao and Junwen Sun contributed equally to this work.}
\affiliation{Key Laboratory for Magnetism and Magnetic Materials of the Ministry of Education, Lanzhou University, Lanzhou, 730000, China}
\author{Junwen Sun}
\thanks{Yu Miao and Junwen Sun contributed equally to this work.}
\affiliation{Department of Physics, The Hong Kong University of Science and Technology, 
Clear Water Bay, Kowloon, Hong Kong, China}
\affiliation{Department of Physics, Nanjing Normal  University, Nanjing, China}
\author{Cunxu Gao}
\affiliation{Key Laboratory for Magnetism and Magnetic Materials of the Ministry of Education, Lanzhou University, Lanzhou, 730000, China}
\author{Desheng Xue}
\email{xueds@lzu.edu.cn}
\affiliation{Key Laboratory for Magnetism and Magnetic Materials of the Ministry of Education, Lanzhou University, Lanzhou, 730000, China}
\author{X. R. Wang}
\email{phxwan@ust.hk }
\affiliation{Department of Physics, The Hong Kong University of Science and Technology, 
	Clear Water Bay, Kowloon, Hong Kong, China}
\affiliation{HKUST Shenzhen Research Institute, Shenzhen, 518057, China}
	
\begin{abstract}
A theory of anisotropic magnetoresistance (AMR) and planar Hall effect (PHE) in single cubic crystals and its experimental verifications are presented for the current in the (001) plane. In contrast to the general belief that AMR and PHE in single crystals are highly sensitive to many internal and external effects and have no universal features, the theory predicts universal angular dependencies of longitudinal and transverse resistivity and various characteristics when magnetization rotates in the (001) plane, the plane perpendicular to the current, and the plane containing the current and [001] direction. The universal angular dependencies are verified by the experiments on Fe$_{30}$Co$_{70}$ single cubic crystal film. The findings provide new avenues for fundamental research and applications of AMR and PHE, because single crystals offer advantages over polycrystalline materials for band structure and crystallographic orientation engineering.
\end{abstract}
	
\maketitle
Anisotropic magnetoresistance (AMR) is a well-known phenomenon that was first discovered in 1856 by Lord Kelvin \cite{Kelvin}. AMR refers to the longitudinal resistance in current direction, which depends on magnetization direction, while the related planar Hall effect (PHE) is the transverse resistance in the Hall geometry. As opposed to an ordinary Hall effect where magnetic field is perpendicular to the Hall plane, the PHE measurement with magnetic field in the Hall plane. AMR and PHE have been extensively studied in magnetic polycrystalline materials \cite{ McGuire1,McGuire2,Ohandley,smit} and single crystals \cite{ Wisniewski,sin-cry-9, gerrit,sin-cry-10,kelly} and their complete understanding is a problem that has persisted for more than 150 years in the field of magnetism.

The universal angular dependencies of AMR and PHE in magnetic polycrystalline materials are well-known \cite{CFJ,book1,book2,Tsunoda1}, which says $\rho_{xx}(\alpha)=\rho_0+A_0\cos^2\alpha$ and $\rho_{xy}(\alpha)=(A_0/2)\sin2\alpha$, where $\alpha$ is the angle between the magnetization and current \cite{Tsunoda2,Jungwirth,yin1,yin2,yin3}. For the AMR and PHE in single crystals, there are also many studies \cite{sin-cry-1,sin-cry-2,sin-cry-3,shi-jing,sin-cry-4,sin-cry-5,sin-cry-6,sin-cry-7,sin-cry-8} that show complicated behaviours \cite{sin-cry-9,sin-cry-10}. However, despite the known roles of spin-orbit interaction, spin-dependent scatterings, and electron interactions with crystallographic directions in AMR, PHE, and extraordinary galvanomagnetic effects \cite{book1,book2}, no universal angular dependencies of AMR and PHE in single crystals have been found to date.

The study of AMR and PHE in single crystals is important for several reasons. First, it should deepen our understanding of the fundamental physics of magnetoresistance, including the role of crystal symmetry and electronic structure. In single crystals, AMR and PHE are directional \cite{my3}, because electronic structures are different along different crystallographic directions, leading to different electron scattering and different group velocities. Second, it enables the development of new materials with tailored magnetic and electronic properties, which can be useful for applications in spintronics such as magnetic recording and sensing \cite{book1,book2}. The in-plane AMR in single crystals has demonstrated higher-order symmetry \cite{sin-cry-9} and phase-shift \cite{fe-jmmm} beyond polycrystalline materials, which may serve as an opportunity for discovering new effects.

In this study, the theory based on vector order parameters for AMR and PHE in single cubic crystals are presented. Through the transport measurements on Fe$_{30}$Co$_{70}$ single cubic crystal film when the current is in the (001) plane with the magnetization rotated in the (001) plane, the plane perpendicular to the current, and the plane containing the current and the [001] direction, the universal angular dependencies of longitudinal and transverse resistivity are verified. We find that only 8 parameters are needed to describe all longitudinal and transverse resistivity curves below the 4th order. We also predict several characteristics of the AMR and PHE, such as the transverse resistivity with current along [100] and [110] directions is identical when magnetization rotating in the above three planes except for the (001) plane. Our results provide new insights of AMR and PHE in single crystal besides magnetization scattering.

In ferromagnetic single crystals, the scattering of electrons is related to crystallographic directions, which can be characterized by three crystalline axis $\vec n_1$,  $\vec n_2$, $\vec n_3$, and the magnetization $\vec M$ whose magnitude is a constant and direction is along $\vec m$. In the linear response region, the electric field $\vec E$ in response to an applied current density $\vec J$ in a crystal must be 
\begin{equation}
\begin{aligned}
\vec E=\tensor{\rho} (\vec m, \vec n_1, \vec n_2, \vec n_3) \vec J,
\label{eq1}
\end{aligned}
\end{equation}
where $\tensor\rho (\vec m, \vec n_1, \vec n_2, \vec n_3) $ is a Cartesian tensor of rank 2. 
Although the tensor values depend on microscopic properties of the crystal and parameters that defines its thermodynamic state, tensor $\tensor\rho$ can be constructed only by $\vec m$,  $\vec n_1$,  $\vec n_2$, and $\vec n_3$. There are ten 
possible Cartesian tensors: $\vec m\vec m$, $\vec n_1\vec n_1$, $\vec n_2\vec n_2$, $\vec n_3\vec n_3$, $\vec m\vec n_1$, $\vec m\vec n_2$, $\vec m\vec n_3$, $\vec n_1\vec n_2$, $\vec n_1\vec n_3$, and $\vec n_2\vec n_3$. Each of them, however, is reducible \cite{Sakurai}, and can be decomposed into the direct sum of a scalar, a vector, and a traceless symmetric tensor. Then, it is possible to construct seven vectors and ten traceless symmetric tensors of ranks 2: $\vec m$, $\vec n_1$, $\vec n_2$, $\vec n_3$, $\vec m\times\vec n_i$, $\vec m \vec m-1/3$, $\vec m \vec n_i+\vec n_i\vec m-2\vec m\cdot\vec n_i/3$, and $\vec n_i\vec n_j+\vec n_j\vec n_i-2\vec n_j\cdot\vec n_j/3$ ($i,j=1,2,3$). Thus, $\tensor\rho$ should be the linear combination of 17 direction-dependent terms together with a scalar term. The electric field $\vec E$ induced by $\vec J$, after grouping similar terms, must take the following most generic form  
\begin{equation}
\begin{aligned}
&\vec E=\rho_0 \vec J+ (B_0 \vec m + \sum_{i=1}^{3}B_i\vec n_i+\sum_{i=1}^{3}B_{3+i}\vec m\times\vec n_i) \times  \vec J\\
&+\sum_{i=1}^{3}A_{i}[(\vec J\cdot \vec m)\vec n_i+(\vec J\cdot \vec n_i)\vec m]+\sum_{i=1}^{3}A_{i+3}(\vec J\cdot \vec n_i)\vec n_i\\
&+A_7 [(\vec J\cdot \vec n_1)\vec n_2+(\vec J\cdot \vec n_2)\vec n_1]+ A_8[(\vec J\cdot \vec n_1)\vec n_3+(\vec J\cdot \vec n_3)\vec n_1]\\
&+A_9[(\vec J\cdot \vec n_2)\vec n_3+(\vec J\cdot \vec n_3)\vec n_2]+A_0(\vec J\cdot \vec m)\vec m ,
\end{aligned}
\label{eq2}
\end{equation}
where $\rho_0$, $A_k$ ($k=0, \ldots, 9$), and $B_l$ ($l=0, \ldots, 6$) are parameters that are determined by the extrinsic and intrinsic properties of a sample such as the temperature, disorders, and the band structures. Of course, these parameters can, in principle, depend on the scalars constructed from $\vec m$ and $\vec n_i$. Among them, only $\vec m\cdot\vec n_i\equiv m_i$ ($i=1,2,3$) can introduce the anisotropic effect. For crystals with reciprocity, $\vec E$ should be the same under $\vec n_i\rightarrow -\vec n_i$ transformations. Thus, $\rho_0$, $A_0$, $A_4$,  $A_5$,  $A_6$, and $B_0$ must be even functions of $m_i$ ($i=1,2,3$). $A_i$, $B_i$, and $B_{i+3}$ must be odd in $m_i$ and even in $m_{j\neq i}$ ($i=1,2,3$) while $A_7$ is odd in $m_1$ and $m_2$, and even in $m_3$. For example, $\rho_0=\sum_{p,q,r}\rho_{0pqr}m_1^{2p}m_2^{2q}m_3^{2r} $ and $A_1=\sum_{p,q,r}A_{1pqr}m_1^{2p+1}m_2^{2q}m_3^{2r}$.
% and $A_7=\sum_{p,q,r}A_{7pqr}m_1^{2p+1}m_2^{2q+1}m_3^{2r}$,
Expansion coefficients $\rho_{0pqr}$ and $A_{1pqr}$ measure the $2(p+q+r)$-th order and $2(p+q+r)+1$-th order interaction strengths of electrons with magnetization and crystal order parameters, respectively, because $A_1$-term contains already one  $\vec m$. Similar expansions can be done for other $A$'s and $B$'s, see Supplementary Information. Because magnetic interactions are usually weak, we shall keep our analysis below the 4th order in the most cases. 

%=====================================================================
\begin{figure*}[t]
\includegraphics*[width=16cm]{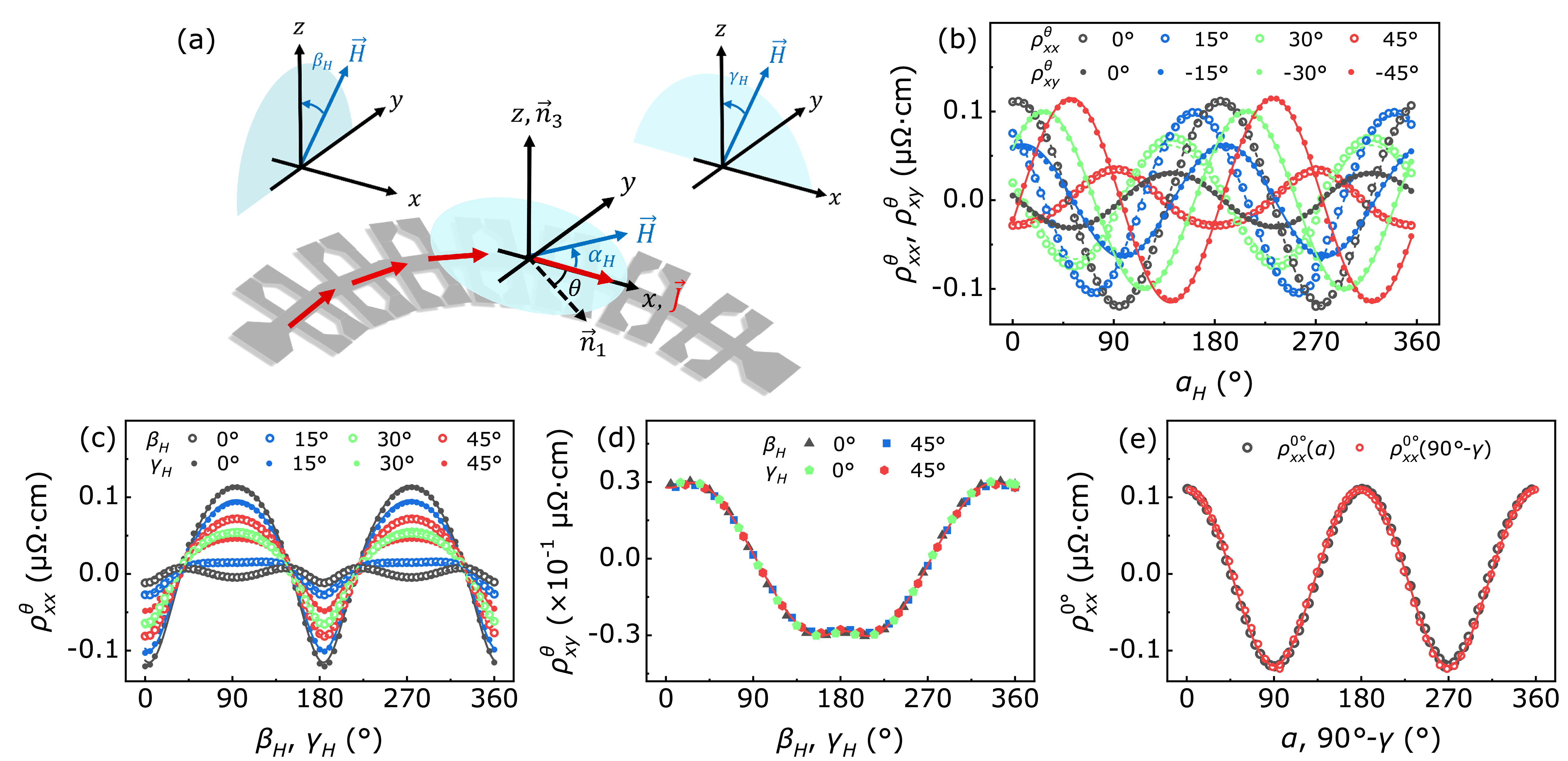}
\caption{The longitudinal and transverse resistivity of Fe$_{30}$Co$_{70}$ single cubic crystal film under 6 T field. (a) The schematics of experimental set-up. Current $\vec J$ is along the $x$-axis in the (001) plane. $\theta$ is the angle between $\vec J$ and $\vec n_1$, and $z$-axis is along the [001] direction. $\alpha_H $ is the angle between magnetic field $\vec H$ and $x$-axis when $\vec H$ is in the (001) plane. $\beta_H$ and $\gamma_H$ are the angles between $\vec H$ and $z$-axis when $\vec H$ is in the $yz$- and $zx$-planes, respectively. (b) $\rho_{xx}^\theta(\alpha_H)$ (open circles) and $\rho_{xy}^\theta(\alpha_H)$ (filled circles) for $\theta=0^\circ, \pm15^\circ, \pm30^\circ$, and $\pm45^\circ$. The dotted (solid) lines are the fitting curves by Eq. (\ref{Supple}) with $\alpha=\alpha_H $. (c) $\rho_{xx}^\theta(\beta_H)$ (open circles) and $\rho_{xx}^\theta(\gamma_H)$ (filled circles) for $\theta=0^\circ, 15^\circ, 30^\circ$, and $45^\circ$. The dotted (solid) lines are the fitting curves by Eq. (\ref{Supple}) with $\beta(\gamma)=\beta_H(\gamma_H)+\delta$. $\delta$ is the angle of magnetization deviated from magnetic field. (d) $\rho_{xy}^\theta(\beta_H)$ (black and blue circles) and $\rho_{xy}^\theta(\gamma_H)$ (green and red circles) for $\theta=0^\circ$ and $45^\circ$. The solid lines are the fitting curves by Eqs. (\ref{0degree}) and (\ref{45degree}) with $\beta(\gamma)=\beta_H(\gamma_H)+\delta$. To display experimental data clearly, only one data point is shown for every four data points collected. (e) $\rho_{xx}^{0^\circ}(\alpha)$ (open black circles) and $\rho_{xx}^{0^\circ}(90^\circ-\gamma)$ (open red circles). They are overlapped with each other and agree with Eq. (\ref{0degree}) (solid line).}
\label{fig1}
\end{figure*}
%=====================================================================

Equation (\ref{eq2}) is the most general electric field response of a crystal to an external current. Among all possible physical quantities, AMR and PHE of a given crystal can be obtained directly from it. In the absence of $\vec n_i$ such as polycrystalline or amorphous magnets, Eq. (\ref{eq2}) reduces to the well-known generalized Ohm's law of polycrystalline materials \cite{yin1,yin2} with only $\rho_0$, $B_0$, and $A_0$ terms. $B_0$-term is the usual anomalous Hall effect, and $A_0$-term is the AMR and PHE for ferromagnetic polycrystalline. If current $\vec J$ is defined as the $x$-axis and the Hall bar is in the $xy$-plane throughout this study, the longitudinal and transverse resistivity are $\rho_{xx}=\rho_0+A_0 \cos^2 \alpha$ and $\rho_{xy}=B_0m_z+(A_0/2)\sin 2\alpha$, where $\alpha$ is the angle between $\vec m$ and $\vec J$. Obviously, $\rho_0$ is the longitudinal resistivity when $\vec J$ is perpendicular to $\vec m$ and $B_0$ is the anomalous Hall coefficient. $A_0$ is the amplitude of the conventional AMR and PHE that is typically a few percent of $\rho_0$. Interestingly, the tensor analysis leads to the famous Einstein gravitation field theory \cite{Carmeli}. The analysis has also been successfully used to predict anomalous spin Hall effects \cite{xrw1,xrw2,xrw3} and unusual AMR in bilayers \cite{xrw4}.

In order to see the implications of Eq. (\ref{eq2}), we apply it to cubic crystals. In this study, a widely used configuration in experiments is considered, where the (001) plane lies on the $xy$-plane, the $z$-axis is along the [001] direction, and $\vec n_1$,  $\vec n_2$, and $\vec n_3$ are equivalent and mutually orthogonal with each other corresponding to [100], [010], and [001] directions. According to Eq. (\ref{eq2}), the longitudinal and transverse resistivity are $\rho_{xx}^\theta \equiv \vec E\cdot \hat x/J=\rho_0 +A_0m_x^2+\sum_{i=1}^{2}(2A_{i}m_xn_{ix}+A_{i+3}n_{ix}^2)+A_7 n_{1x}n_{2x}$ and $\rho_{xy}^\theta \equiv \vec E\cdot \hat y/J=B_0m_z+B_3+B_4 (m_xn_{1y}-m_yn_{1x})+B_5(m_xn_{2y}-m_yn_{2x})+A_0m_xm_y+ \sum_{i=1}^{2}[A_{i}(m_xn_{iy}+m_yn_{ix})+A_{i+3}n_{ix}n_{iy}]+A_7(n_{1x}n_{2y}+n_{1y}n_{2x})$\textcolor{blue}, where $\theta$ is the angle between the [100] direction and the $x$-axis.

Following the convention in literature, we define $\alpha$ as the angle between $\vec m$ and $\vec J$ when $\vec m$ rotates in the $xy$-plane, $\beta$ and $\gamma$ as the angles between $\vec m$ and $z$-axis when $\vec m$ rotates in the $yz$- and $zx$-planes, respectively, as illustrated in Fig. \ref{fig1}(a). After some tedious algebras, as shown in the Supplementary Information, the angular dependencies of $\rho_{xx}^{\theta}$ and $\rho_{xy}^{\theta}$, with terms not higher than $m_i^4$, are
\begin{equation}
\begin{aligned}
&\rho_{xx}^{\theta}(\alpha)=\rho_1 \cos 2\alpha + \rho_2 \cos (2\alpha+4\theta)+\rho_3 \cos (4\alpha+4\theta),\\
&\rho_{xx}^{\theta}(\beta)=(\mu_1-\mu_1 \cos 4\theta)\cos 2\beta +(\rho_4+\rho_5 \cos 4\theta)\cos 4\beta,\\
&\rho_{xx}^{\theta}(\gamma)= (\mu_2+\mu_3 \cos 4\theta)\cos 2\gamma+(\mu_4+\mu_5 \cos 4\theta)\cos 4\gamma,\\
&\rho_{xy}^{\theta}(\alpha)=\rho_1 \sin 2\alpha -\rho_2 \sin (2\alpha+4\theta) -\rho_6 \sin(4\alpha+4\theta),\\
&\rho_{xy}^{\theta}(\beta)= \rho_7 \cos \beta+\sin 4\theta(\mu_6 \cos 2\beta +\mu_7  \cos 4\beta)+\rho_8 \cos 3\beta,\\ 
&\rho_{xy}^{\theta}(\gamma)=\rho_7 \cos \gamma+\sin 4\theta(\mu_{8}\cos 2\gamma +\mu_{9}\cos 4\gamma)+\rho_8 \cos 3\gamma,
\end{aligned}
\label{Supple}
\end{equation}
where $\rho_i$ ($i=1,2,\ldots,8$) are the only independent constants which depend on material parameter. $\mu_i$ 
($i=1,2,\ldots,9$) are linear combinations of $\rho_i$ ($i=1,2,\ldots,8$) and are
$\mu_1 = -(\rho_2 -\rho_3 )/2$, $\mu_2 = \mu_1 -\rho_1$, 
$\mu_3=\mu_1 - \rho_3$, $\mu_4 = 3\rho_3/4+\rho_5$, $\mu_5=\rho_3-\mu_4$, $\mu_6 = -(\rho_2 -\rho_6 )/2,$
$\mu_7= \rho_3/8 -\rho_5+ \rho_6/8$,
$\mu_8 = \mu_6+\rho_2$, $\mu_9 =-\rho_6/4-\mu_7$. We have also removed the angular independent background resistance such that averaged $\rho_{xx}^{\theta}(\alpha)$ with respect to $\alpha$ is zero.

For $\vec J$ along [100] ($\theta =0^\circ$) and [110] ($\theta =45^\circ$ and equivalent to $\theta =-45^\circ$ or $[\bar 1\bar1 0]$), we have 
\begin{equation}
\begin{aligned}
&\rho_{xx}^{0^\circ}(\alpha)=(\rho_1+\rho_2)\cos 2\alpha + \rho_3\cos 4\alpha,\\
&\rho_{xx}^{0^\circ}(\beta)=(\rho_4+\rho_5) \cos 4\beta ,\\
&\rho_{xx}^{0^\circ}(\gamma)=-(\rho_1+\rho_2)\cos 2\gamma + \rho_3\cos 4\gamma ,\\
&\rho_{xy}^{0^\circ}(\alpha)=(\rho_1-\rho_2)\sin 2\alpha-\rho_6\sin 4\alpha ,  \\
&\rho_{xy}^{0^\circ}(\beta)=\rho_7\cos\beta +\rho_8\cos 3\beta , \\ 
&\rho_{xy}^{0^\circ}(\gamma)=\rho_7\cos\gamma +\rho_8\cos 3\gamma, 
\end{aligned}
\label{0degree}
\end{equation}
and 
\begin{equation}
\begin{aligned}
&\rho_{xx}^{45^\circ}(\alpha)=(\rho_1-\rho_2)\cos 2\alpha - \rho_3\cos 4\alpha, \\
&\rho_{xx}^{45^\circ}(\beta)=(\rho_3-\rho_2)\cos 2\beta+(\rho_4-\rho_5)\cos 4\beta, \\
&\rho_{xx}^{45^\circ}(\gamma)=(\rho_3-\rho_1)\cos 2\gamma+(\frac{1}{2}\rho_3+2\rho_5) \cos 4\gamma, \\
&\rho_{xy}^{45^\circ}(\alpha)=(\rho_1+\rho_2)\sin 2\alpha + \rho_6\sin 4\alpha,  \\
&\rho_{xy}^{45^\circ}(\beta)=\rho_7\cos\beta +\rho_8\cos 3\beta,   \\ 
&\rho_{xy}^{45^\circ}(\gamma)=\rho_7\cos\gamma +\rho_8\cos 3\gamma.
\end{aligned}
\label{45degree}
\end{equation}

Interestingly, there are several characteristics according to Eqs. (\ref{0degree}) and (\ref{45degree}). 1) The amplitude of 2-fold in $\rho_{xx}^{0^\circ(45^\circ)}(\alpha)$ is equal to that in $\rho_{xy}^{45^\circ(0^\circ)}(\alpha)$, and the amplitude of 4-fold in $\rho_{xx(xy)}^{0^\circ}(\alpha)$ and $\rho_{xx(xy)}^{45^\circ}(\alpha)$ are always the same. 2) $\rho_{xx}^{0^\circ}(\beta)$ have no 2-fold symmetry but strictly 4-fold symmetry. 3) The results of $\rho_{xy}^{0^\circ}(\beta)$, $\rho_{xy}^{0^\circ}(\gamma)$, $\rho_{xy}^{45^\circ}(\beta)$, and $\rho_{xy}^{45^\circ}(\gamma)$ which only have a 1-fold and a 3-fold term are identical, and $\rho_{xx}^{0^\circ}(90^\circ-\gamma)$ is identical to $\rho_{xx}^{0^\circ}(\alpha)$. Furthermore, the sum of amplitudes of 2-fold terms in $\rho_{xx}^{0^\circ(45^\circ)}(\alpha)$ and $\rho_{xx}^{0^\circ(45^\circ)}(\gamma)$ equals to that in $\rho_{xx}^{0^\circ(45^\circ)}(\beta)$. Other relationships among the angular dependencies of longitudinal and transverse resistivity also exist and can be used to test the theory. By inspection, early experiments of AMR and PHE in References \cite{shi-jing} and \cite{sin-cry-9} for (Ga,Mn)As and Co$_x$Fe$_{1-x}$ single cubic crystal films along such two special angles   agree with our theory.

To verify the theory presented above, we measured angular dependences of $\rho_{xx}^\theta$ and $\rho_{xy}^\theta$ of Fe$_{30}$Co$_{70}$ single crystal film. A  19-nm-thick Fe$_{30}$Co$_{70}$ single crystal film was grown on MgO(001) substrate at room temperature by molecular beam epitaxy. The single crystal sample is patterned into Hall bars along different crystallographic direction using photo-lithography and ion beam etching as schematically shown in Fig. \ref{fig1}(a). In one batch, we fabricated Hall bars along $\theta=0^\circ, \pm15^\circ, \pm30^\circ$, and $\pm45^\circ$ with size of $1000~\mu$m$\times 50~\mu$m. Both the longitudinal and transverse resistivity $\rho_{xx}^{\theta}$ and $\rho_{xy}^{\theta}$ are measured using the four-probe method. The results for current in the (001) plane of our Fe$_{30}$Co$_{70}$ film are plotted in Fig. \ref{fig1}(b)-\ref{fig1}(d). The symbols are experimental data (after subtracting the background resistances, and $\rho_{xy}^{\theta}$ divided by a coefficient of 1.19 due to the effect of the finite electrode size for the Hall measurement \cite{sin-cry-9,fe-jmmm}). In order to compare the experimental results with the theoretical prediction, $\alpha$, $\beta$, and $\gamma$ of magnetization should be derived from the corresponding angles $\alpha_H$, $\beta_H$, and $\gamma_H$ of magnetic field which can be determined experimentally. A 6 T magnetic field is applied to ensure the magnetization close to the direction of field. Then $\alpha\simeq \alpha_H$ because of the magnitude of in-plane magnetocrystalline anisotropy field is two orders of magnitude smaller than the applied field strength. $\beta\simeq \beta_H+\delta$ and $\gamma\simeq \gamma_H+\delta$ since the out-of-plane shape anisotropy field is about 2 T which is not much smaller than 6 T. The angle $\delta$ of magnetization deviated from magnetic field can be expressed as \cite{my1,my2}, 
\begin{equation}
\delta(\beta_H)=\frac{\sin2\beta_H}{2(H/H_K-\cos2\beta_H)},
\label{conversion}
\end{equation}
where $ H $ is the magnitude of magnetic field, and $ H_K $ is the anisotropy field. Equation (\ref{conversion}) is also applicable to $\gamma_H$. The dashed and solid lines in Fig. \ref{fig1}(b)-\ref{fig1}(d) are fitting curves by Eq. (\ref{Supple}) with 8 fitting constants given in  Tab. \ref{tab1} after converting $\alpha$, $\beta$, and $\gamma$ to  $\alpha_H$, $\beta_H$, and $\gamma_H$. 

%=====================================================================
\begin{table}[h]
		\centering
		\caption{The fitting parameters in Eq. (\ref{Supple}) for Fig. \ref{fig1}(b)-\ref{fig1}(d).}
		\begin{ruledtabular}
		\begin{tabular}{lcccc}
		    $ \times 10^{-2} $&$ \rho_1 $&$ \rho_2 $&$ \rho_3 $&$ \rho_4 $\\\hline
		    $ \mathrm{\mu\Omega\cdot cm } $&4.185&7.337&-0.446&-0.653\\
		    &$ \rho_5 $&$ \rho_6 $&$ \rho_7 $&$ \rho_8 $\\\hline
			$ \mathrm{\mu\Omega\cdot cm } $&-0.145&-0.104&3.452&-0.650\
		\end{tabular}
		\end{ruledtabular} 
		\label{tab1}
\end{table}
%=====================================================================

The characteristics of single cubic crystals from the theory can be verified experimentally. Figure \ref{fig1}(b) shows the AMR and PHE in the $ xy$-plane. The amplitude of AMR gradually decreases while that of PHE increases with current applied from $ \theta=0^{\circ}$ to $ \theta=45^{\circ} $. The amplitudes of $\rho_{xy}^{45^\circ(0^\circ)}(\alpha)$ and $\rho_{xx}^{0^\circ(45^\circ)}(\alpha)$ are the same as repdicted. Figure \ref{fig1}(c) shows the AMR in the $ yz-$ and the $zx$-plane. Only a net 4-fold symmetry appear in $\rho_{xx}^{0^\circ}(\beta)$, in total agreement with the theory. Figure \ref{fig1}(d) shows the transverse resistivity in the $ yz-$ and the $ zx$-plane. All four $\rho_{xy}^\theta(\beta)$ and $\rho_{xy}^\theta(\gamma)$ for $\theta=0^\circ$ and $45^\circ$ are coincident. Figure \ref{fig1}(e) shows the AMR in terms of $\alpha$ and $(90^\circ-\gamma)$ using the angle conversions mentioned above. $\rho_{xx}^{0^\circ}(90^\circ-\gamma)$ and $\rho_{xx}^{0^\circ}(\alpha)$ are the same as predicted by our theory. Our experimental measurements support umambigurosly all characteristics summarized early. 

%=====================================================================
\begin{figure}[t]
\centering
\includegraphics*[width=8cm]{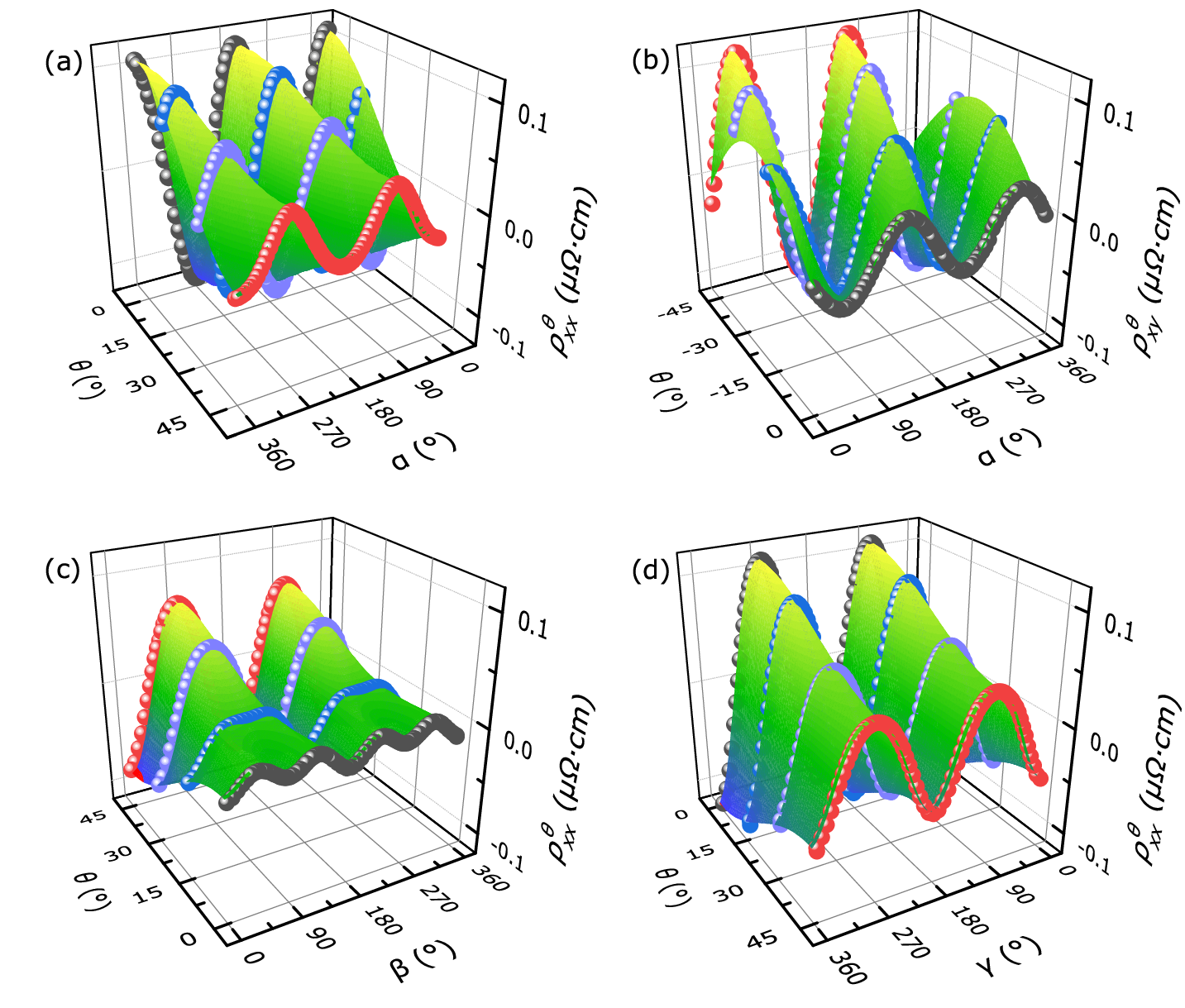}
\caption{Three-dimensional plots of AMR and PHE as functions of current and magnetization directions when the current is in the (001) plane. Symbols are experimental data of (a,c,d) AMR and (b) PHE in terms of $\theta$ and $\alpha$ (a,b), $\theta$ and $\beta$ (c), and $\theta$ and $\gamma$ (d). The space curved surfaces are Eq. (\ref{Supple}) with $\rho_i $ ($i=1,2,\ldots, 8$) given in Tab. \ref{tab1}.}
\label{fig2}
\end{figure}
%=====================================================================

To have a better picture of how $\rho_{xx}$ and $\rho_{xy}$ vary with the current direction ($\theta$) and the direction of $\vec m$, we convert $\alpha_H$, $\beta_H$, and $\gamma_H$ to $\alpha$, $\beta$, and $\gamma$, and plot $\rho_{xx}^\theta$ and $\rho_{xy}^\theta$ as functions of $\theta$ and $\alpha$, or $\beta$, or $\gamma$ in Fig. \ref{fig2}. The three dimensional surfaces are the theoretical formula of Eq. (\ref{Supple}) with parameters given in Tab. \ref{tab1}. The beautiful agreements of experiments and theory in the 3D plots are a strong testimony of correctness of the theory, meaning clearly that only 8 independent parameters can indeed describe all longitudinal and transverse resistivity curves.

\begin{figure}[h]
\includegraphics[width=8cm]{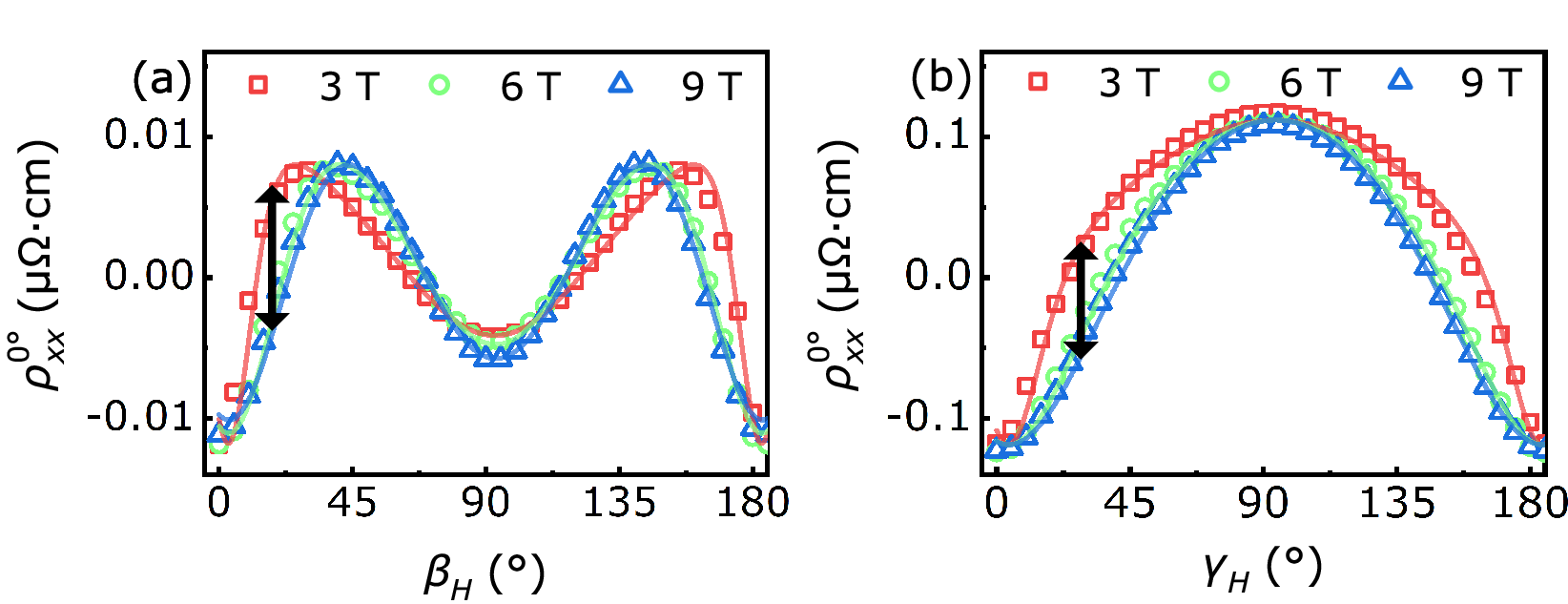}
\caption{AMR with current along $ \theta=0^\circ $ under different magnetic fields. $\rho_{xx}^{0^\circ}$ vs. $\beta_H$ (a) and $\gamma_H$ (b) for magnetic fields of 3 T, 6 T, and 9 T. Symbols are the experimental data, and the solid lines are Eq. (\ref{0degree}). Black bidirectional arrow is used to indicate differences between the curves. }
\label{fig3}
\end{figure}

To test how good our field and magnetization direction correction is, we also measure AMR at different fields with current applied along  the [100] crystallographic direction. Figure \ref{fig3} is $\rho_{xx}^{0^\circ}(\beta_H)$ (a) and $\rho_{xx}^{0^\circ}(\gamma_H)$ (b) for field at 3 T (red squares), 6 T (green circles), and 9 T (blue triangles). Although the AMR curves are significantly different with increasing fields, especially around $\beta_H(\gamma_H)=22.5^\circ$, the results can also be well fitted by Eqs. (\ref{0degree}) with the same parameters in Tab. \ref{tab1}, revealing the field-independence of the 8 parameters as suggested by the theory. Of course, the angles in Eqs. (\ref{Supple}) are converted to $\beta_H$ and $\gamma_H$ by Eq. (\ref{conversion}).

%=====================================================================
%=====================================================================

It must be mentioned that there are fundamental differences between current tensor analysis and the symmetry consideration \cite{McGuire1,Birss,Rout} widely used to understand AMR in single crystals. Although the symmetry consideration is not wrong, it does not reveal universal angular-dependencies of AMR and PHE and possible identities. One obtains different results presented here when the symmetry analysis is applied on a cubic crystal \cite{sin-cry-10}. In fact, it cannot even recover the universal behaviour of AMR and PHE in polycrystalline without extra inputs \cite{sin-cry-10}. One recent progress connects AMR with the magnetization-dependent band structure near the Fermi level \cite{sin-cry-9,gerrit}. These density functional calculations are consistent with our theory although they cannot identify universal angular dependencies and characteristics in AMR and PHE. 

In summary, a generic AMR and PHE theory for single cubic crystals are presented. Only 8 intrinsic parameters are needed to describe angular dependencies of magnetoresistance. A set of characteristics among $\rho_{xx}^{\theta}$ and $\rho_{xy}^{\theta}$ are predicted for current along the [100] and [110] directions and the magnetization rotating in the $xy$-, $xz$-, and $yz$-planes. The predictions are beautifully verified by the experiments on Fe$_{30}$Co$_{70}$ single cubic crystal film. We believe that the long-standing issue of universal angular dependencies of AMR and PHE in single cubic crystals is resolved.

\begin{acknowledgments}
This work is supported by the National Key 
Research and Development Program of China (No. 
2020YFA0309600), the NSFC Grant (No. 91963201, 
12374122, 12174163, and 12074157) and Hong Kong RGC Grants (No. 16300522, 16300523, and 16302321). 
\end{acknowledgments}

\end{document}